\documentclass[11 pt,twocolumn,amsmath,amssymb]{article}

\usepackage[utf8]{inputenc}
\usepackage[english]{babel}
\usepackage[left=1.5cm, right=1.5cm, top=1.785cm, bottom=2.0cm]{geometry}
\usepackage[usenames,dvipsnames]{xcolor}
\usepackage{color}
\usepackage[most]{tcolorbox}
\usepackage{caption}
\usepackage{subcaption}
\usepackage[percent]{overpic}
\usepackage{graphicx}
\usepackage{array}
\usepackage{authblk}

\usepackage{cite}

\begin{document}

\title{\bf{Spectrally stable thermal emitters enabled by material-based high-impedance surfaces}}
\author[1]{David Navajas}
\author[1]{Jos\'e M. P\'erez-Escudero}
\author[1]{I\~{n}igo Liberal}

\affil[1]{\it{Department of Electrical, Electronic and Communications Engineering, Institute of Smart Cities (ISC), Public University of Navarre (UPNA), 31006 Pamplona, Spain.}}
\date{} % clear date
\twocolumn[
  \begin{@twocolumnfalse}
    \maketitle
    \begin{abstract}
Radiative thermal engineering with subwavelength metallic bodies is a key element for heat and energy management applications, communication and sensing. Here, we numerically and experimentally demonstrate metallic thermal emitters with narrowband but extremely stable emission spectra, whose resonant frequency does not shift with changes on the nanofilm thickness, the angle of observation and/or polarization. Our devices are based on epsilon-near-zero (ENZ) substrates acting as material-based high-impedance substrates. They do not require from complex nanofabrication processes, thus being compatible with large-area and low-cost applications.
    \end{abstract}
  \end{@twocolumnfalse}  \vspace{0.6cm}
  ]
\section{Introduction}

The emission and absorption of infrared radiation by subwavelength metallic bodies is the basic operating principle behind a variety of technologies including thermal camouflage \cite{Shahsafi2019,Kats2014}, heat-management \cite{Naqavi2018,Li2015}, optoelectronic devices  \cite{Krayer2019}, thermophotovoltaics \cite{Fleming2002all}, novel light sources \cite{Greffet2002,Ilic2016}, radiative cooling \cite{Fan2022}, photochemistry \cite{Mukherjee2013hot} and bioengineering \cite{Shi2015}. The advantage of concentrating the absorption/emission on small metallic bodies is that they minimize the volume where the process takes place, leading to faster and stronger modulations with temperature. Increasing the speed and dynamic range of the modulation expedites the performance of communication devices \cite{Sakat2018,Wojszvzyk2021incandescent}, the impact of nonequilibrium dynamics \cite{Park2020,Inoue2014}, photo-thermal modulation \cite{Dias2020} and metal-to-insulator transitions \cite{qu2017dynamic,Wan2019optical} empowering negative differential thermal emittance \cite{kats2013vanadium}, temperature-independent emission \cite{shahsafi2019temperature}, and thermal homeostasis \cite{Wu2017thermal,Ono2018self}. Furthermore, concentrating the absorption within a deeply subwavelength region catalyzes basic research on sub-nanometer and atomically-thin metallic materials \cite{Lenert2014,AbdEl-Fattah2019}, and their nonlocal \cite{Mortensen2014} and quantum \cite{Zhu2016quantum} response. 

Conventional approaches to the design of narrowband and partially coherent thermal emitters include the use of structured surfaces supporting leaky modes \cite{Greffet2002}, nanoresonators \cite{Schuller2009optical,Celanovic2005resonant,Andueza2021}, photonic crystals \cite{Luo2004thermal,Laroche2006}, resonant multi-layered structures \cite{Kats2013nanometre,Rensberg2017epsilon,Kumagai2020,Zhao2020}, metasurfaces \cite{Costantini2015plasmonic,Blanchard2022}, transformation optics \cite{Alwakil2017} and angle selective filters \cite{Sakr2017}, to name a few. These different strategies could be adapted to concentrate the absorption/emission within a subwavelength metallic body. However, all these approaches, albeit different, share common properties arising from the fact that they rely on the excitation of resonant modes. 

First, their emission spectrum is polarization- and angle-dependent, following the spatial dispersion of the resonances. This property is convenient for the design of selective emitters, but avoids the design of emitters with a stable emission spectrum. Similarly, the resonant frequency is defined by the geometry of the nanostructure. Therefore, fabrication tolerances modify the spectrum via frequency shifts and inhomogeneous broadening. Finally, the need of fabricating nanostructures capable of supporting specific resonant modes limits their applicability for large-area, large-scale and low-cost thermal emitters.   

Here, we propose the use of epsilon-near-zero (ENZ) substrates, acting as a material-based high-impedance surfaces (HISs), as a pathway to obtain spectrally stable emitters. HISs consist of metamaterial constructs that replicate the boundary conditions of an artificial magnetic conductor (AMC). Thus, HISs enforce a maximum of the electric field tangential to its surface, strengthening the interactions with thin metallic films. HISs have been successfully employed at microwave frequencies to develop electromagnetic absorbers \cite{Tretyakov2003thin,Kern2003genetic}, low-profile antennas \cite{feresidis2005artificial,baracco2008amc}, reconfigurable antennas \cite{Sievenpiper2003two,Costa2008active}, anti-radar surfaces \cite{Paquay2007thin,Cui2014coding} and suppressing surface waves \cite{Sievenpiper1999high}. By contrast, material-based HISs take advantage of the material properties of a dispersive medium to implement a HISs without the need of fabricating a metamaterial construct \cite{Liberal2022vision,Perez2020}. Here, we numerically and experimentally demonstrate that material-based HISs exhibit an unusually stable absorptivity/emissivity spectrum, characterized by a narrowband peak, whose frequency position is independent of the geometry of the metallic film, the angle of observation and polarization.

Our results resonate with previous works highlighting geometry-independent phenomena in ENZ media \cite{Liberal2016zero}, including supercoupling \cite{Silveirinha2006tunneling}, photonic doping \cite{Liberal2017photonic}, geometry-invariant resonators \cite{Liberal2016geometry} and frequency pinning of nanoantenna resonances \cite{kim2016role}. All of these effects are based on wavelength expansion and spatial delocalization in ENZ media. By contrast, here we take advantage of the extreme boundary conditions of ENZ media to induce a stable emission peak in an otherwise nonresonant metallic film, thus opening a new route to geometry-invariant phenomena in ENZ media. We believe that our results introduce a novel class of thermal emitters, and represent an important step forward in the development of practical ENZ-based technologies. 

\section{Theory and numerical results}

\begin{figure}[h!]
 \centering
  \includegraphics[width=1\linewidth]{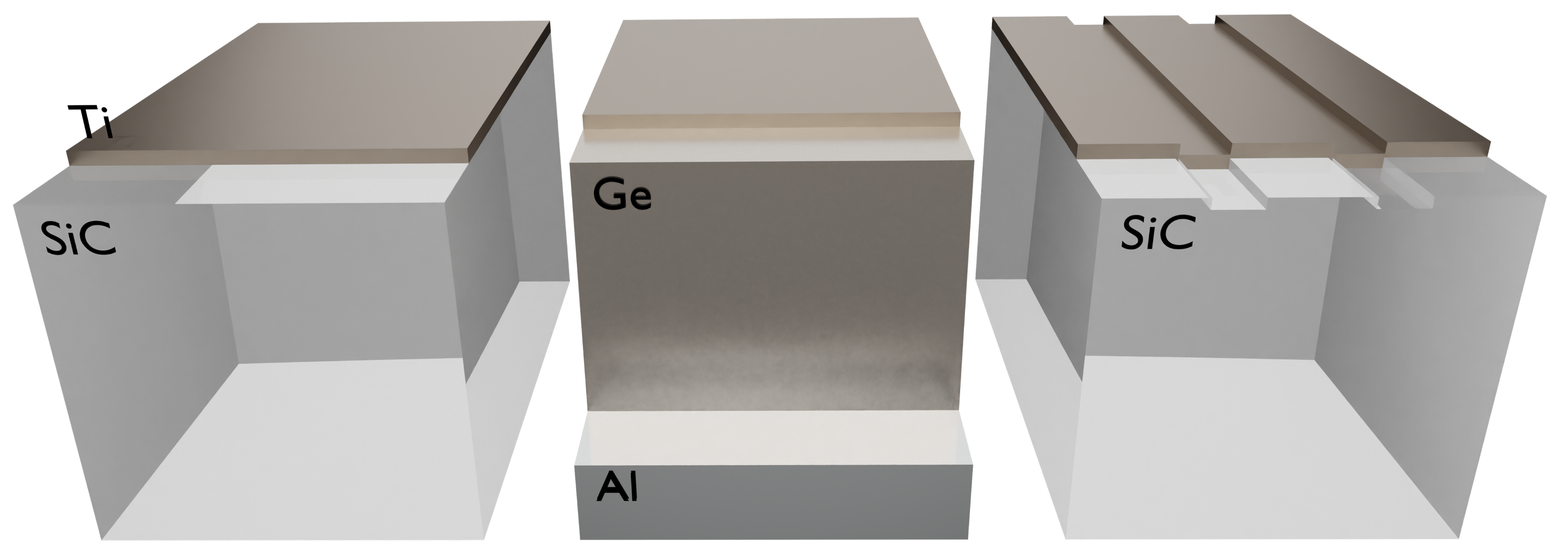}
 \caption{\textbf{Sketch of comparative case studies for enhancing the absorptivity/emissivity of an ultra-thin titanium (Ti) film.} (Left) Material-based high-impedance surface (HIS) based on a silicon carbide (SiC) substrate. (Center) Salisbury screen implemented with a Germanium (Ge) spacer and an aluminum (Al) mirror. (Right) Rectangular corrugations on a SiC substrate acting as a grating supporting leaky modes.}
 \label{fig:configurations}
\end{figure}

We start by numerically investigating the spectral performance of material-based HISs (high-impedance surfaces)  against variations of the geometry, angle of observation and polarization. As a particular material platform, the absorptivity/emissivity of a Titanium (Ti) thin film, deposited on top of a silicon carbide (SiC) substrate is studied. In this doing so, we will find that material-based HISs, implemented with ENZ substrates, exhibit an unusually stable absorptivity/emissivity spectrum. To highlight that this is not how conventional nanostructured thermal emitters behave, we provide the comparative case studies with a Salisbury screen and a grating supporting a leaky mode (sketches of geometries can be seen in Fig.\,\ref{fig:configurations}).

\begin{figure*}[h]
\centering
\begin{tcolorbox}[
  colback=white!5!white,
  colframe=blue!75!black,
  title={\centering \scriptsize{Transversal Magnetic mode (TM)}},sharp corners]
    \begin{subfigure}{0.33\linewidth}
   \begin{overpic}[width=\linewidth]{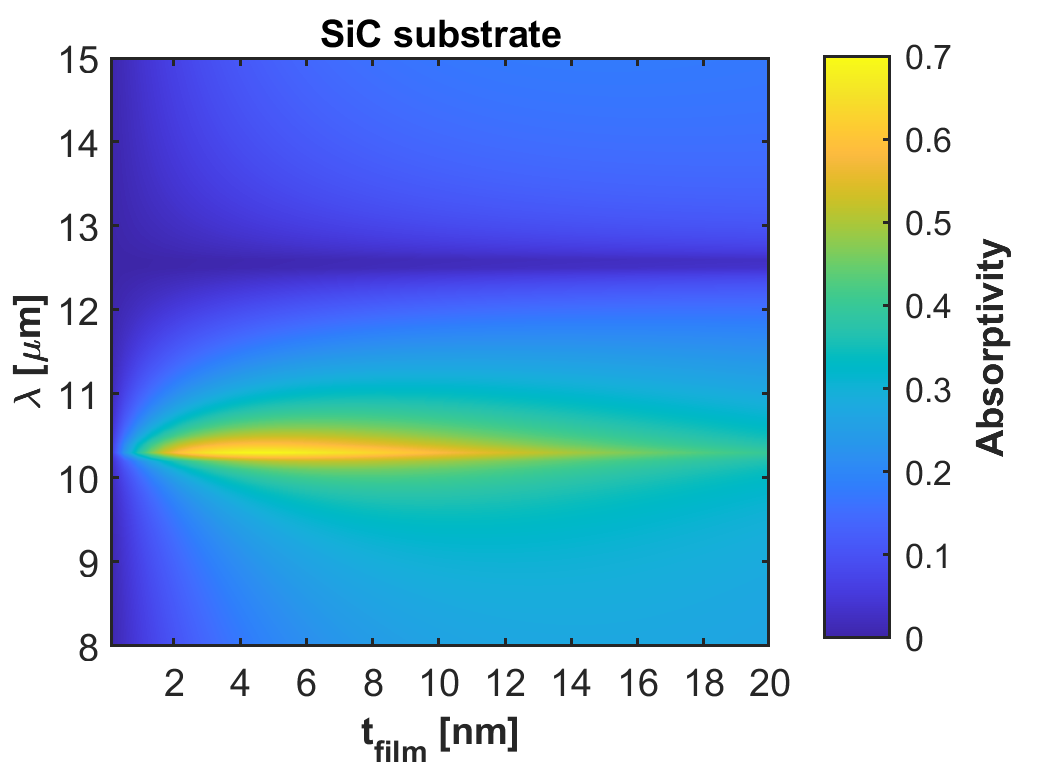}
     \put(47,45){\includegraphics[scale=0.1]{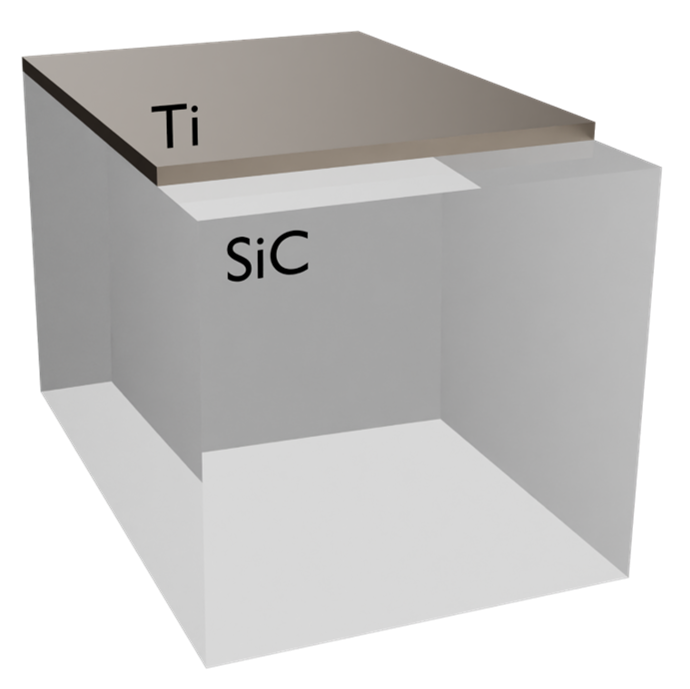}}
  \end{overpic}
    \caption{}
\label{fig:TiSiCTMgrosor}
    \end{subfigure}\hfill
    \begin{subfigure}{0.33\linewidth}
    \begin{overpic}[width=\linewidth]{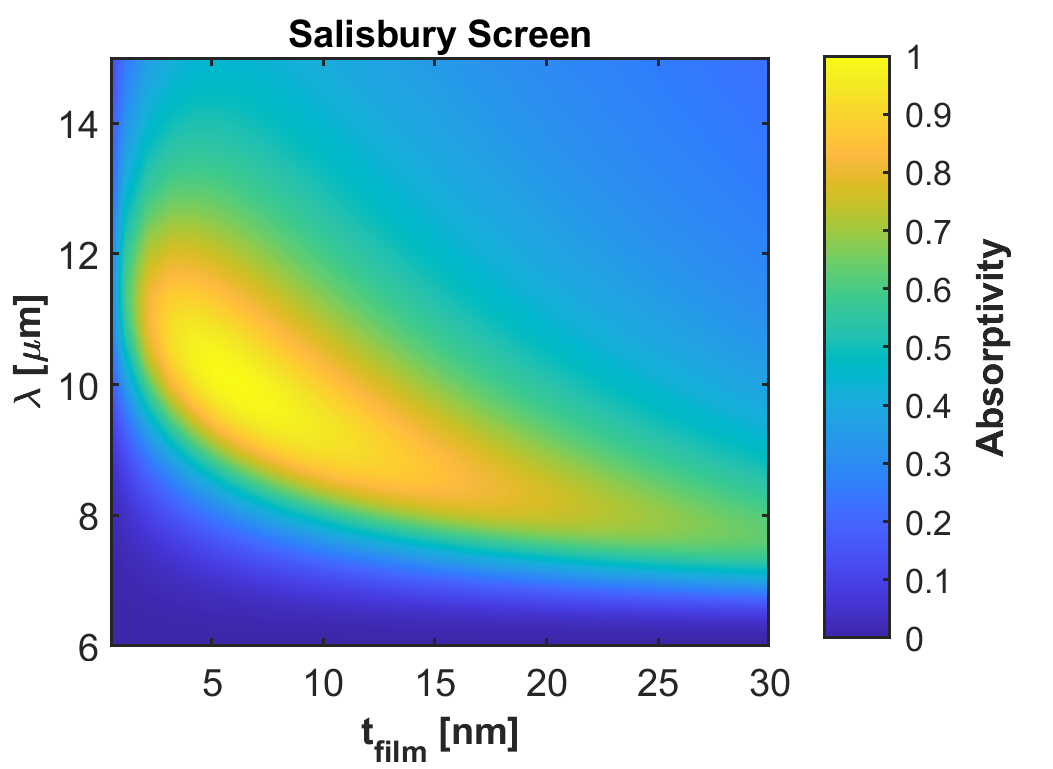}
     \put(47,45){\includegraphics[scale=0.1]{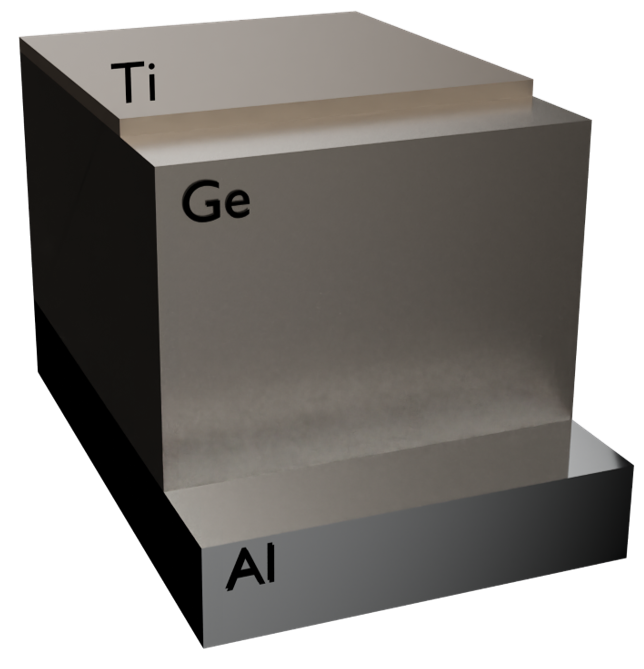}}
  \end{overpic}
    \caption{}
\label{fig:TiGeTMgrosor}
    \end{subfigure}\hfill
    \begin{subfigure}{0.33\linewidth}
    \begin{overpic}[width=\linewidth]{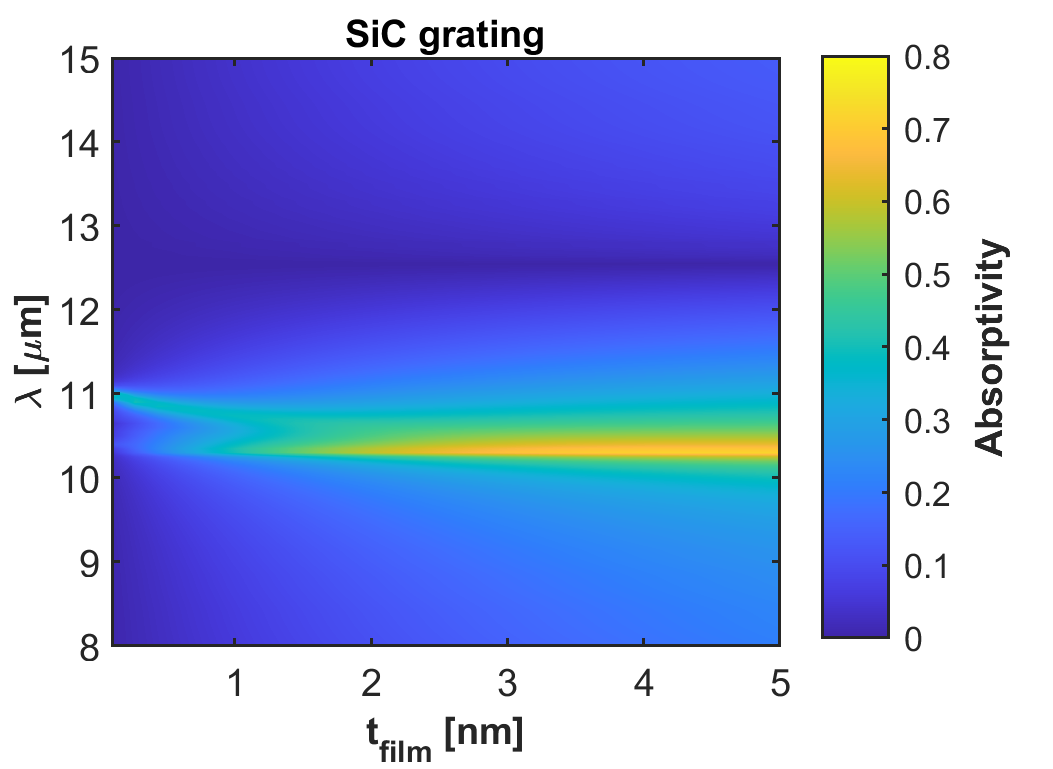}
     \put(47,45){\includegraphics[scale=0.1]{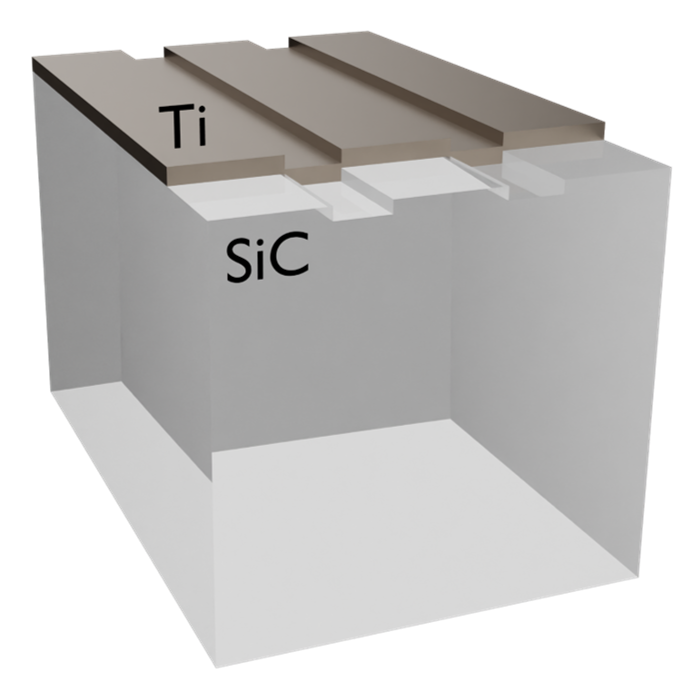}}
  \end{overpic}
    \caption{}
\label{fig:gratingTMgrosor}
    \end{subfigure}
 \end{tcolorbox}
 \begin{tcolorbox}[
  colback=white!5!white,
  colframe=blue!75!black,
  title={\centering \scriptsize{Transversal Electric mode (TE)}},sharp corners]
    \begin{subfigure}{0.33\linewidth}
\includegraphics[width=\linewidth]{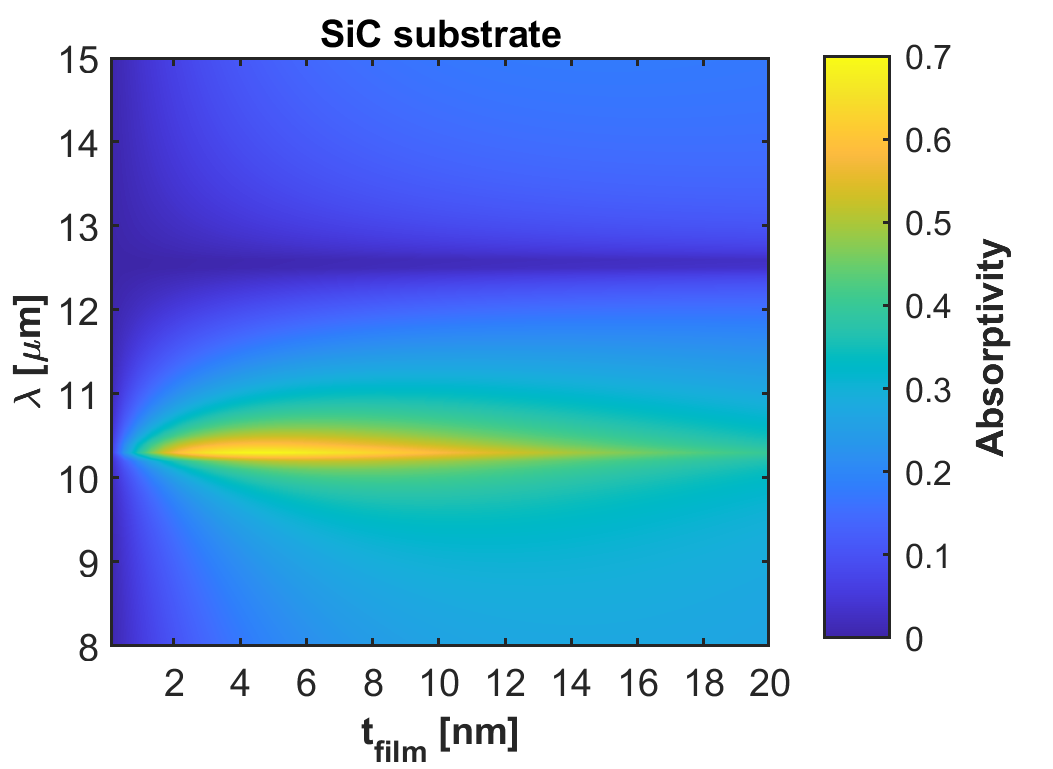} 
    \caption{}
\label{fig:TiSiCTEgrosor}
    \end{subfigure}\hfill
    \begin{subfigure}{0.33\linewidth}
\includegraphics[width=\linewidth]{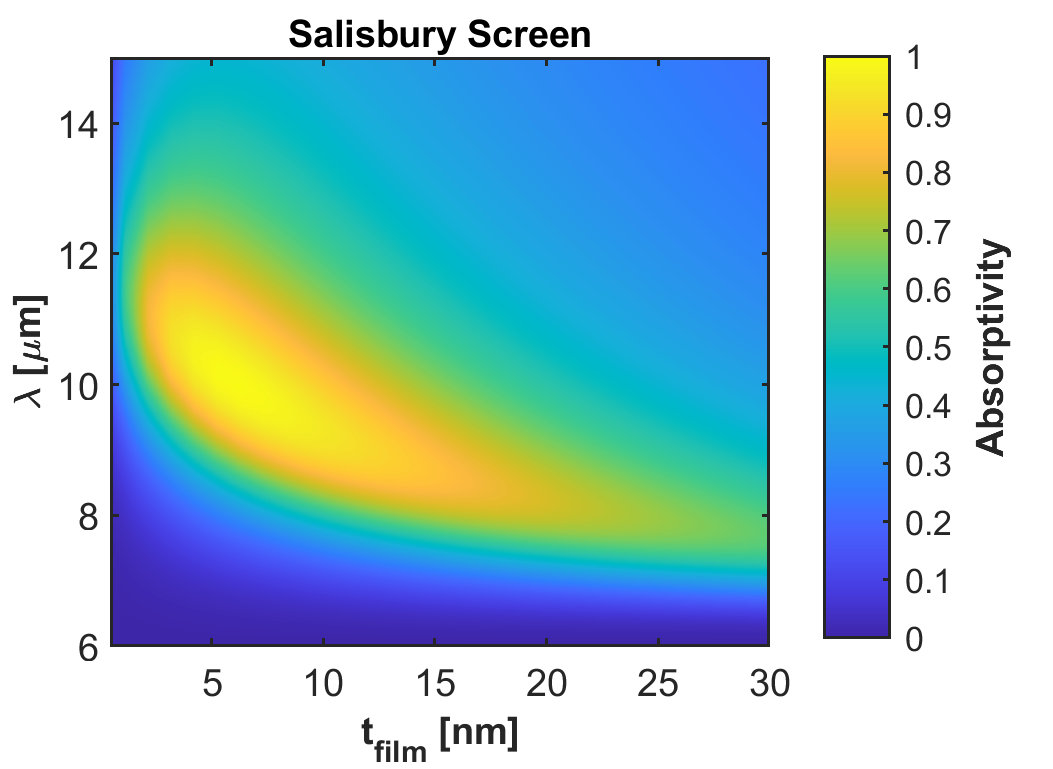}
    \caption{}
\label{fig:TiGeTEgrosor}
    \end{subfigure}\hfill
    \begin{subfigure}{0.33\linewidth}
\includegraphics[width=\linewidth]{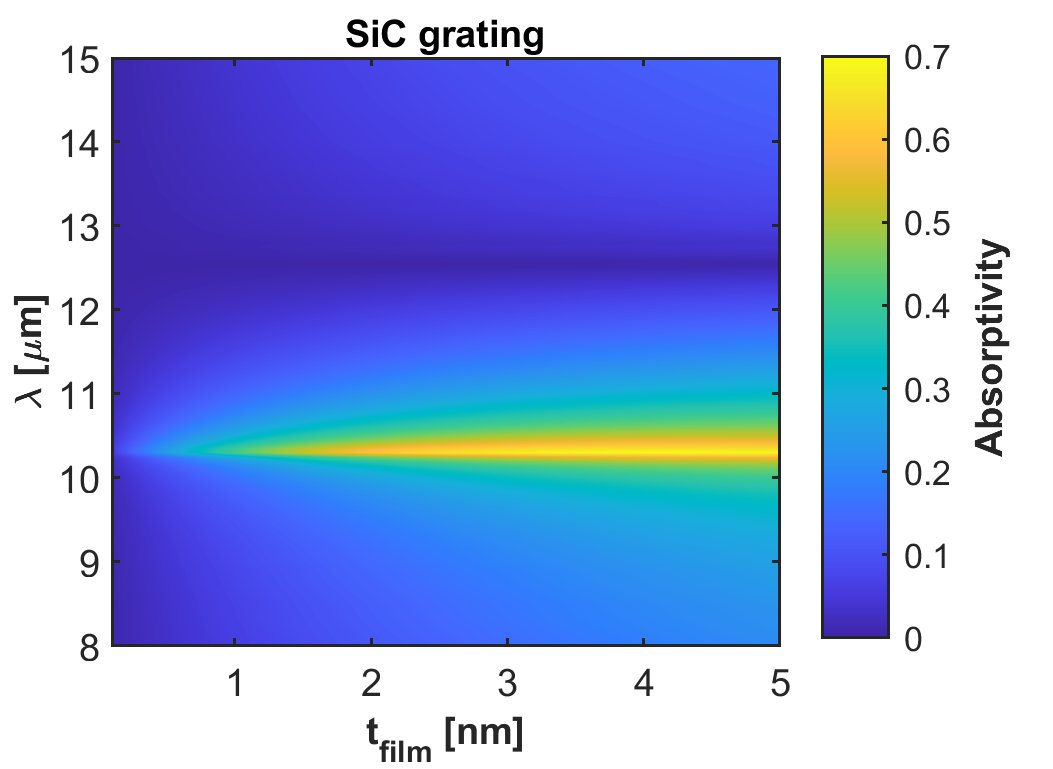}
    \caption{}
\label{fig:gratingTEgrosor}
    \end{subfigure}
  \end{tcolorbox}
\caption{{\bf Impact of the titanium (Ti) film thickness on the absorptivity/emissivity spectrum.} Numerical prediction of the film absorptivity for 
(left column) a Ti thin film on top of a SiC substrate, (center column) a Salisbury screen composed of the Ti film, a germanium (Ge) spacer, and a perfect electric conductor (PEC) mirror, and (right column) a Ti film on top of a SiC grating with rectangular corrugations with a 0.55$\lambda_p$ pitch and a $\lambda_p/40$ depth for $\lambda_p = 11.36 \mu$m. The results are shown for the TM (first row) and TE (second row) polarizations. Insets: Sketch of the geometry.}
    \label{fig:thicknessvariation}
    \end{figure*}

 \begin{figure*}[h]
\centering
\begin{tcolorbox}[
  colback=white!5!white,
  colframe=blue!75!black,
  title={\centering \scriptsize{Transversal Magnetic mode (TM)}},sharp corners]
    \begin{subfigure}{0.33\linewidth}
 \begin{overpic}[width=\linewidth]{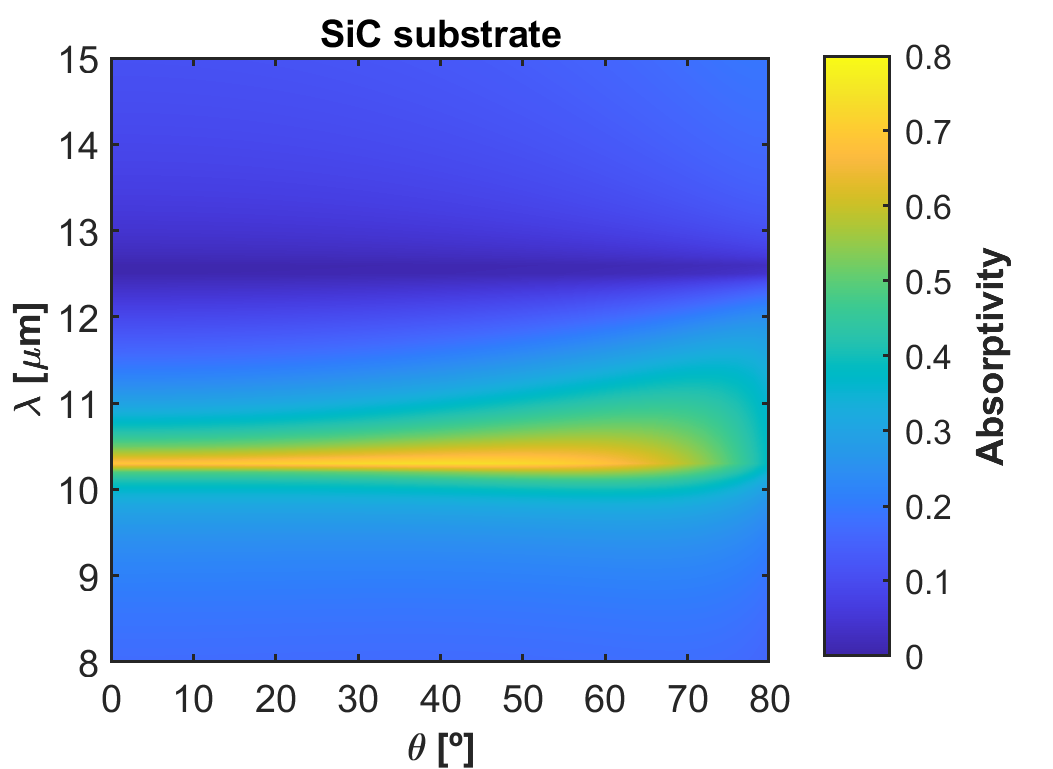}
     \put(15,45){\includegraphics[scale=0.1]{SiC_R.png}}
  \end{overpic}
    \caption{}
\label{fig:2a}
    \end{subfigure}\hfill
    \begin{subfigure}{0.33\linewidth}
\begin{overpic}[width=\linewidth]{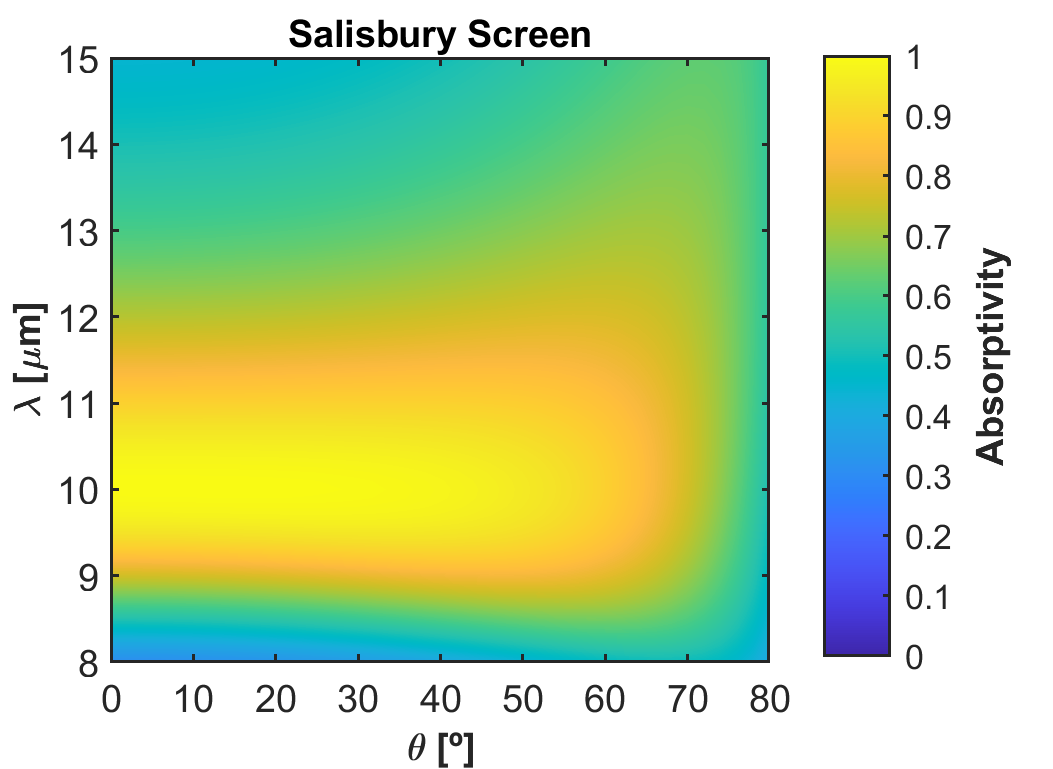}
     \put(15,45){\includegraphics[scale=0.1]{Salisbury_R.png}}
  \end{overpic}
    \caption{}
\label{fig:2b}
    \end{subfigure}\hfill
    \begin{subfigure}{0.33\linewidth}
    \begin{overpic}[width=\linewidth]{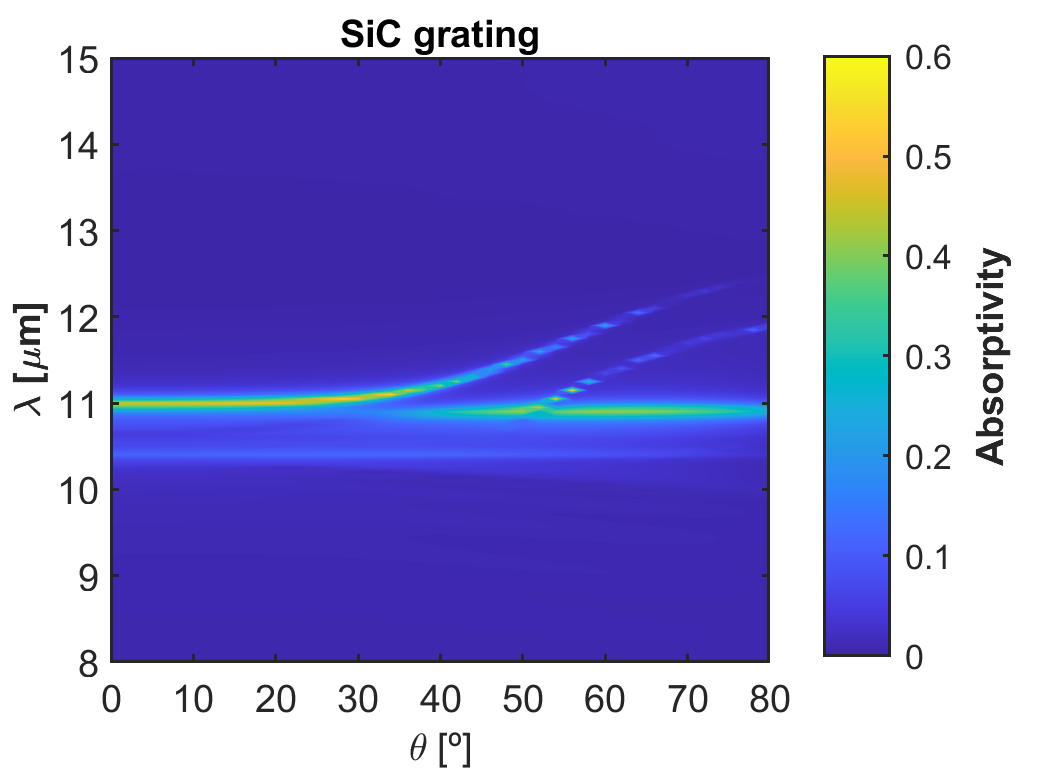}
     \put(15,45){\includegraphics[scale=0.1]{Grating_R.png}}
  \end{overpic}
    \caption{}
\label{fig:2c}
    \end{subfigure}
    \end{tcolorbox}
    \begin{tcolorbox}[
  colback=white!5!white,
  colframe=blue!75!black,
title={\centering \scriptsize{Transversal Electric mode (TE)}},sharp corners]
    \begin{subfigure}{0.33\linewidth}
\includegraphics[width=\linewidth]{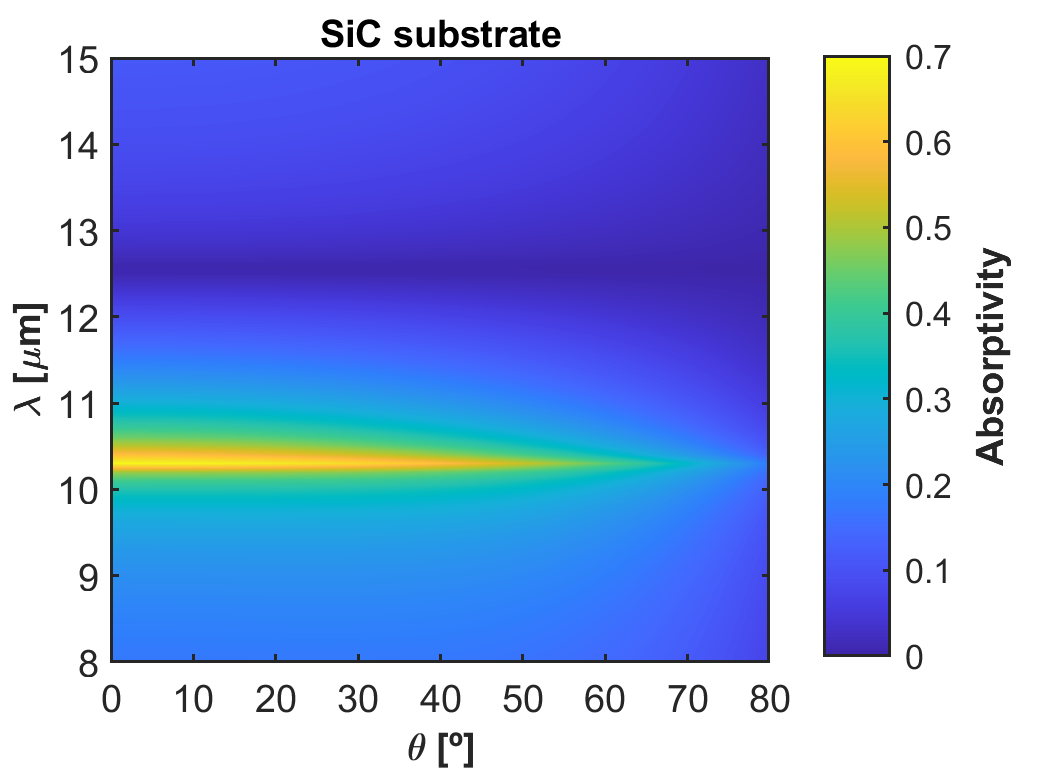} 
    \caption{}
   
\label{fig:2d}
    \end{subfigure}\hfill
    \begin{subfigure}{0.33\linewidth}
\includegraphics[width=\linewidth]{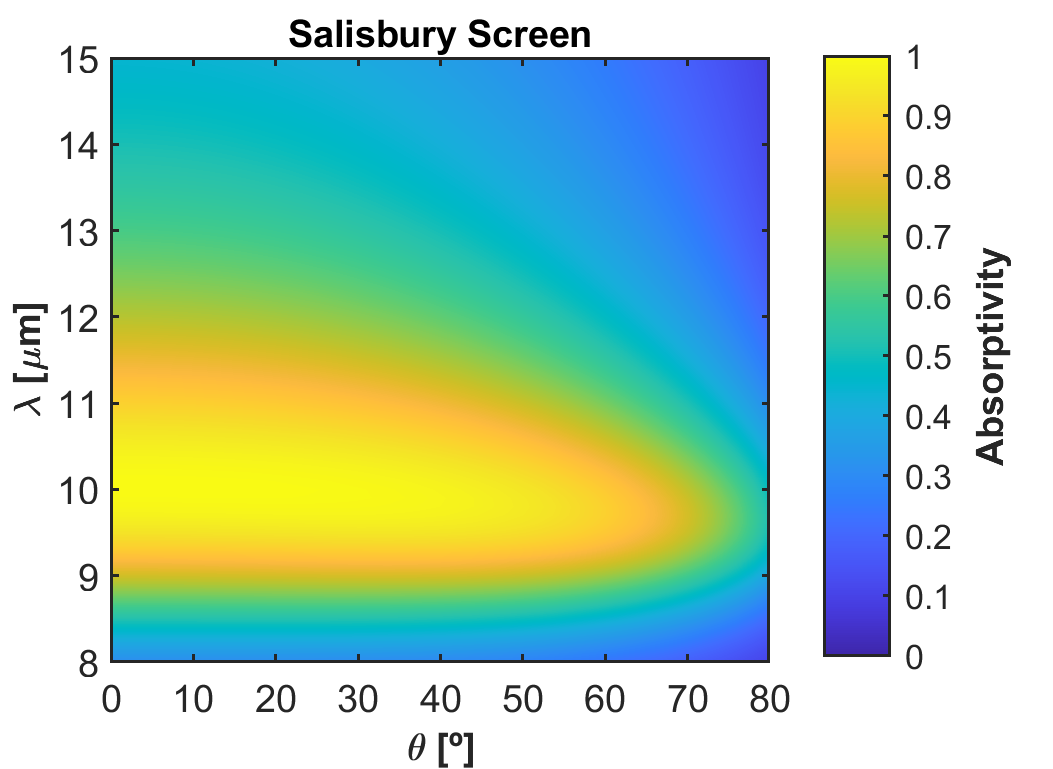}
    \caption{}
\label{fig:2e}
    \end{subfigure}\hfill
    \begin{subfigure}{0.33\linewidth}
\includegraphics[width=\linewidth]{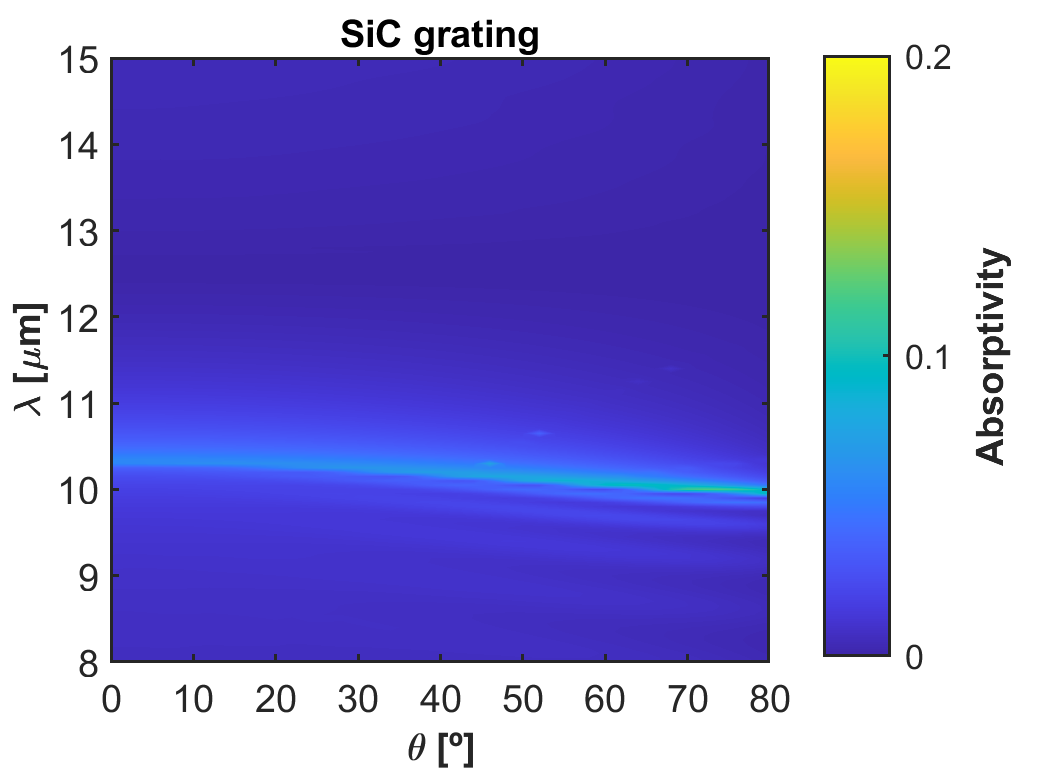}
    \caption{}
\label{fig:2f}
    \end{subfigure}
    \end{tcolorbox}
\caption{{\bf Impact of the observation angle on the absorptivity/emissivity spectrum.} Numerical prediction of the film absorptivity, as a function of the incident angle, for (left column) a Ti thin film of 4 nm thickness on top of a SiC substrate, (center column) a Salisbury screen composed of a Ti film of 6 nm thickness, a germanium (Ge) spacer of 830 nm thickness, and a perfect electric conductor (PEC) mirror, and (right column) a Ti film on top of a SiC grating with rectangular corrugations with a 0.55$\lambda_p$ pitch and a $\lambda_p/40$ depth for $\lambda_p = 11.36 \mu$m. The selected Ti film thicknesses correspond to those maximizing the absorptivity at normal incidence. The results are shown for the TM (first row) and TE (second row) polarizations. Insets: Sketch of the geometry.}
    \label{fig:2}
    \end{figure*}
    
\subsection{Material-based HISs}

Silicon carbide (SiC) is characterized by a relative permittivity with a Lorentzian dispersion profile \cite{Caldwell2015low}, $\varepsilon_{SiC}(\omega) = \varepsilon_{\infty}(\omega^2 - \omega_p^2 + i\omega\omega_c)/(\omega^2 - \omega_0^2 + i\omega\omega_c)$, with normalized medium impedance $Z_{SiC}(\omega)=1/\sqrt{\varepsilon_{SiC}(\omega)}$. The plasma frequency $\omega_p$ is located around a wavelength of $\lambda_p = 10.3\,\mu m$, where SiC behaves as an ENZ medium, $\varepsilon_{SiC}(\omega_p)\rightarrow 0$. Thus, a SiC substrate acts as a material-based HIS, $Z_{SiC}(\omega_p)\gg 1$, that enhances the tangential electric fields on its surface, and strengthens the interaction between electromagnetic fields and ultra-thin metallic films. On the other hand, the resonance frequency $\omega_0$ is located around a wavelength of $\lambda_0 = 12.55\,\mu m$, where the permittivity of SiC acquires very high permittivity values $\left|\varepsilon_{SiC}(\omega_0)\right|\gg 1$. Consequently, near the resonance frequency a SiC substrate acts as a low-impedance surface, $Z_{SiC}(\omega_0)\rightarrow 0$, with a behavior similar to that of metallic mirrors.

Because the high-impedance character of SiC substrates directly arises from its material properties, it avoids the need of complex nanofabrication processes, as it is the case in conventional, metamaterial-based, HISs. Furthermore, here we demonstrate that material-based HISs exhibit a narrowband but extremely stable spectral response. First, the strong frequency dispersion of SiC near its plasma frequency leads to a narrowband absorptivity/emissivity response, enabling for the design of partially coherent thermal emitters without the need of nanostructuring a high quality-factor resonant structure. Similarly, because the substrate impedance sweeps a wide range of values around the plasma frequency, changes in the geometry, angle of observation and/or polarization do not induce significant frequency shifts. In this manner, material-based HISs are characterized by an unusually stable emission spectrum.

We illustrate this point by performing full wave numerical simulations (see Methods) on a titanium (Ti) thin film deposited on top of a SiC substrate. Following the local form of Kirchhoff's law \cite{Greffet2018light}, we numerically calculate thermal emissivity from the metallic film through its absorptivity (see Methods). Figures \ref{fig:TiSiCTMgrosor} and \ref{fig:TiSiCTEgrosor} depict the predicted film absorptivity/emissivity spectra for transversal magnetic (TM) and transversal electric (TE) polarizations, respectively, as a function of the film thickness. It can be concluded from the figure that, for all film thicknesses, the film absorptivity/emissivity is maximized at the plasma frequency, where the substrate acts as a HIS, while it is minimized at the resonance frequency, where the substrate behaves the closest to a metallic mirror. The maximum peak value 69,4\,\% is obtained for a thin-film thickness of approximately 4.5 nm. It represents an enhancement of 2.8 times the maximum value that could be obtained with Ti, standing in free-space (24.63\,\%), and an even better performance as compared to the film deposited on a nondispersive dielectric substrate with $\varepsilon_{r}>1$. Also, we re-emphasize that this figure of merit corresponds to the absorption taking place only within the ultra-thin metallic film. Again, the frequency position of the maximum peak is extremely stable, remaining fixed at $\lambda= 10.3\,\mu$m in the studied range from 1\,nm to 20\,nm thickness. Despite the fact that all studied film thicknesses are deeply subwavelength, films with similar characteristics induce important frequency shifts in conventional thermal emitters, as we will show in the following examples.

The robustness of the emission frequency against variations of the metallic film thickness provides with important technological advantages. For example, accurately controlling the nanometric film thickness over a large area is an important technical challenge. Similarly, producing a high-quality thin-film with a small roughness requires from a carefully controlled fabrication process. Material-based HISs exhibit a stable spectral response against variations of the thin film geometry. Thus, they provide an interesting alternative for large-area, large-scale applications, as well as for low-cost nanofabrication techniques. 

Due to the rotational symmetry of the system, the response at normal incidence is the same for both TE and TM polarizations in Fig.\,\ref{fig:TiSiCTMgrosor} and \ref{fig:TiSiCTEgrosor}, and for all film thicknesses. However, this symmetry is broken at oblique incidence, in general leading to variations on the emission spectrum as a function of the observation angle. Figure\,\ref{fig:2a} and \ref{fig:2d} depict the numerical prediction of the angular response for the TE and TM polarizations, respectively, for the film of 4\,nm thickness that maximized the film absorptivity/emissivity at normal incidence. Normal incidence corresponds to $\theta$=0\textdegree and grazing incidence corresponds to $\theta$=90\textdegree. Our numerical simulations reveal that changing the observation angle does not induce a frequency shift in the film absorptivity/emissivity. In other words, the metal film will exhibit a peak of thermal emission at a fixed frequency, independently of the direction along which it is measured. Because no geometrical resonance is used to induce a narrow linewidth, one can simultaneously observe angularly-independent and narrowband thermal emission.

\subsection{Multilayered structures: Salisbury screen}

It can be concluded from our numerical simulations that material-based HISs enable thermal emitters with a spectrally stable response. Importantly, we remark that this is not how conventional partially coherent emitters behave. To illustrate this point, we next analyze the spectral response of Salisbury screens. The Salisbury screen is a popular absorber in microwave engineering \cite{Fante1988reflection,Munk2005frequency} that recently has gained popularity in the design of thermal emitters \cite{Bakan2018,Wang2015}. A Salisbury screen basically consists of a three layer structure, formed by a resistive layer, placed on a $\lambda/4$ spacer on top of a metallic mirror. 

The bandwidth and angular dependence of a Salisbury screen are inversely proportional to the dielectric permittivity. Therefore, in order to provide the toughest comparison with the material-based HIS, we consider a Salisbury screen implemented with a germanium (Ge) spacer \cite{Li1980}, which exhibits a high dielectric constant ($\varepsilon_{Ge}\simeq 16$) at infrared frequencies. Salisbury screens implemented with spacers with a lower dielectric constant feature wider bandwidths and a stronger angular variability (see Supplementary Fig.1).

Because the operating principle of the Salisbury screen critically depends on the resonant behavior induced by the spacer, it is well-known that the emission frequency of the Salisbury screen is very sensitive against variations of the spacer thickness, thus requiring from precise nanofabrication processes. Next, it is shown that the emission frequency of a Salisbury screen also critically depends on the thickness of the metallic layer. To clarify this point, Fig.\,\ref{fig:TiGeTMgrosor} and \ref{fig:TiGeTEgrosor} depict the numerical prediction of the absorptivity/emissivity for a Salisbury screen constructed with an ideal perfect electric conductor (PEC) mirror, a Ge spacer of 730 nm thickness, and a Ti metallic film of different thicknesses, for TE and TM polarizations, respectively. It can be concluded from the figure that the peak emission wavelength presents a strong dependence on the film thickness. Specifically, the peak emission wavelength varies from 11.8\,$\mu m$ for films with a vanishing thickness, to a peak emission wavelength of $\lambda\simeq 7.7 \mu m$ for large thicknesses. 

The large variation effect is due to the internal inductance of the Ti layer. In particular, for vanishing film thickness, peak emission occurs at the wavelength where the condition $tan(\omega/c\sqrt{\varepsilon_{spacer}}\,t_{spacer})=\infty$ is satisfied. It corresponds to the $\lambda/4$ resonance of a conventional Salisbury screen, composed of a purely resistive layer. However, as the thickness of the film increases, the impact of its internal inductance can no longer be neglected, shifting the resonance to higher frequencies (See Supplementary Fig.2). For large film thicknesses, the peak emission occurs at the wavelength $\lambda\simeq 8 \mu m$, decreasing more and more when reaching the  $tan(\omega/c\sqrt{\varepsilon_{spacer}}\,t_{spacer})=0$ condition. The approximated thickness for which the highest peak of absorption is obtained is 6.2 nm with a peak value of 99.91\%. The value is fairly superior to that obtained with the SiC substrate, where losses limited the range of the substrate impedance. By contrast, we modeled Ge with an ideally lossless permittivity, and the metallic mirror was modeled with an ideal PEC. Moreover, the superior peak performance comes at the cost of a broader and less stable spectral performance. The fact that the Salisbury screen peak emission wavelength depends strongly not only with the space thickness, but also with nanometric variations of the metallic film thickness, might reduce its applicability for large-area and/or low cost applications.

Figures \ref{fig:2b} and \ref{fig:2e} depict the absortivity/emissivity of the Salisbury screen with Ti film of 6 nm thickness, for TM and TE polarization, respectively, and as a function of the incident angle. The figures show that, despite featuring a wider bandwidth, the peak wavelength of the Salisbury screen shifts as a function of the observation angle. For the TM polarization, the peak shifts from $\lambda=10\,\mu$m at normal incidence, to $\lambda=10.5\,\mu$m at $\theta = 80$\,\textdegree. For the TE polarization, the peak shifts from $\lambda=10\,\mu$m at normal incidence, to $\lambda=9.7\,\mu$m at $\theta = 80$\,\textdegree. Here, we re-emphasize that a Ge-based Salisbury screen was selected as a competitive test, and that Salisbury screens based on spacers with a lower permittivity will lead to a larger angular variability. 

\subsection{Nanostructured surfaces: gratings supporting leaky modes}

Another approach towards the design of partially-coherent thermal emitters is to provide the substrate with a structure at the nanoscale, either by sculpting nanoresonators, nanoantennas and/or gratings. All these cases are characterized by reduced mode volumes and strong coupling to electromagnetic fields, albeit at the cost of more accurate nanofabrication processes. As an example of a nanostructured thermal emitters, we consider the spectral performance of metallic thin-films on top of gratings. Gratings are characterized by a periodic modulation of the surface, which modifies the dispersion properties of surfaces modes, reducing its propagation constant and thus allowing the coupling to free-space modes \cite{Greffet2002}. Surface modes provide strong confinement of the field near the surface of the substrate, and they can strongly couple to a metallic thin-film deposited on top of the substrate. To illustrate this point we consider a grating consisting of rectangular rods etched on a SiC substrate, with a 0.55$\lambda_p$ pitch and a $\lambda_p/40$ depth for $\lambda_p = 11.36 \mu$m.

Figures \ref{fig:gratingTMgrosor} and \ref{fig:gratingTEgrosor} depict the numerically predicted variation of the film absorptivity/emissivity as a function of the film thickness, for both TM and TE polarizations, respectively. In this case, the system no longer has rotational symmetry, and it presents a different behavior for TM and TE fields. In fact, the grating mode only effectively couples to TM radiation, and the response for the TE polarization is very similar to that of the bare SiC substrate. For the TM polarization, the film absortivity/emissivity presents a strong dependence of the film thickness. In fact, the results presented in figure \ref{fig:gratingTMgrosor} reveal that peak absortivity/emissivity range from $\lambda = 11\,\mu$m for films with a vanishing thickness, to $\lambda = 10.3\,\mu$m for a thickness of only 3 nm. For thicknesses greater than 3 nm, the grating mode merges with the response of the SiC substrate acting as a material-based HIS at the ENZ frequency. In the region where leaky mode is well-defined, a local maximum film absorptivity/emissivity of 47.95\% is found for a film thickness of 0.1\,nm. Strong coupling of SiC nanoresonators to dielectric sub-nanometer films has been experimentally demonstrated in previous works \cite{Berte2018sub}. Consequently, our numerical simulations show that sub-nanometer metallic films already exhibit strong absorption efficiencies, and a large variability on the emission spectrum. The strong variability of the absorptivity/emissivity of the metallic film arises from the strong electric field confinement provided by resonant nanostructures. While this property is very convenient for the design of high-sensitivity sensors, it directly leads to inhomogeneous broadening for small variations of the film geometrical properties. In fact, our numerical simulations predict that small variations of the film thickness, comparable the dimension of typical roughness, induce a significant variation on the peak wavelength and/or might provoke the disappearance of the leaky mode.

\begin{figure}
 \centering
\includegraphics[width=\linewidth]{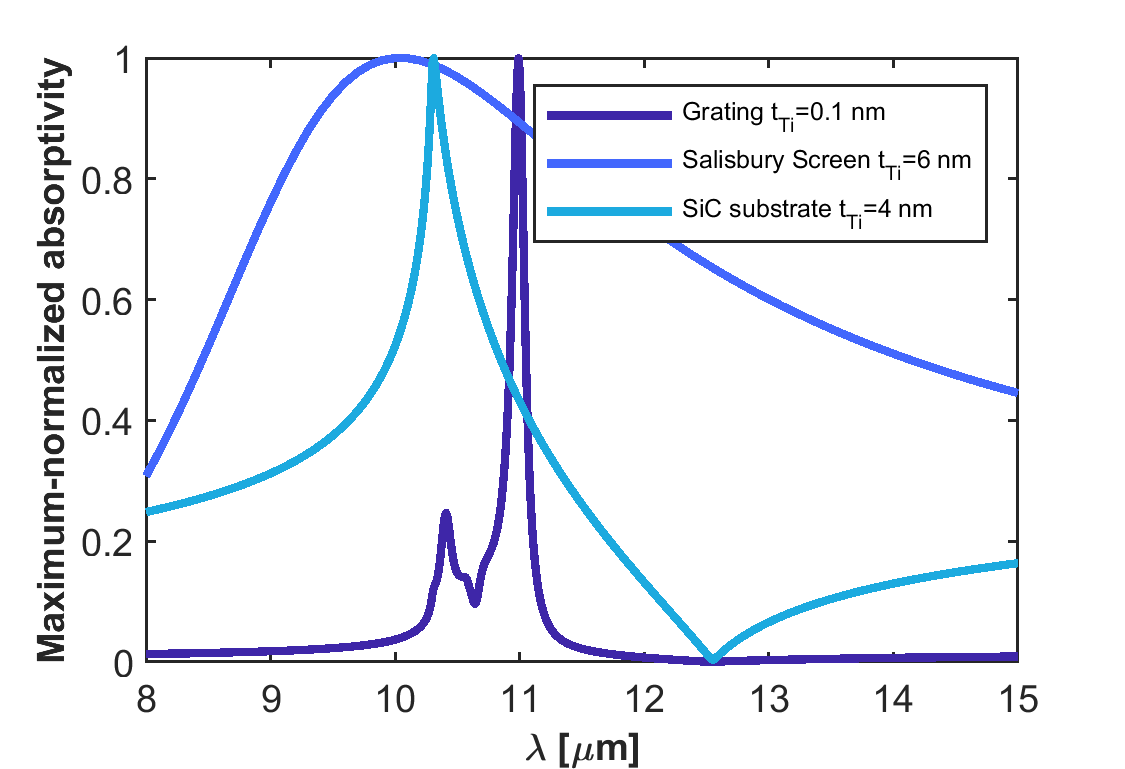}
 \caption{{\bf Comparison of bandwidth performance}. Numerical prediction of the film absorptivity spectra, normalized to its maximum value, for a Ti thin film of 4 nm thickness on top of SiC substrate, a Salisbury screen composed of a Ti film of 6 nm thickness, a germanium (Ge) spacer of 730 nm thickness, and a perfect electric conductor (PEC) mirror, and a Ti film on top of a SiC grating with rectangular corrugations with a 0.55$\lambda_p$ pitch and a $\lambda_p/40$ depth for $\lambda_p = 11.36 \mu$m.}
 \label{fig:anchobanda}
\end{figure}

Figures \ref{fig:2c} and \ref{fig:2f} depict the numerical prediction of the film absorptivity/emissivity for a Ti film thickness of 0.1 nm as a function of the observation angle, for both TM and TE polarizations, respectively. Again, the grating mode can only be observed for the TM polarization, while a response similar to that of the bare substrate is observed for the TE polarization. For the TM polarization, the wavelength of the peak absorptivity/emissivity shifts from $\lambda = 11\,\mu$m at normal incidence to $\lambda = 12.45\,\mu$m for $\theta = 80$\,\textdegree. The strong dependence of the film absorptivity/emissivity with the angle of observation is a common property of gratings, where the peak wavelength follows the dispersion of the leaky mode. For this reason, thermal emitters based on gratings supporting leaky modes exhibit a different spectrum as a function of the observation angle.

To finalize our comparative study, Fig.\,\ref{fig:anchobanda} reports a comparison of the bandwidth performance for the three different classes of thermal emitters studied above, for the Ti film thicknesses that maximized the peak absorptivity/emissivity on each case. It can be concluded from the figure that material-based HISs provide a bandwidth much smaller than that obtained with the competitive Ge-based Salisbury screen. %(see also Supplementary Fig.\,7).
The figure also reveals that material-based HISs provide a bandwidth comparable to that obtained with the grating supporting a leaky mode, despite the fact that the fabrication of a high quality factor resonator was not needed. 

In conclusion, ENZ substrates acting as material-based HISs enable the design of narrowband thermal emitters whose spectral response is unusually stable against variations of the geometry, observation angle and polarization. In addition, such properties are obtained even without the need of complex nanofabrication processes.

\section{Experimental Results} %------------------------

\begin{figure}[t]
 \centering
 \begin{subfigure}{\linewidth}
\includegraphics[width=\linewidth]{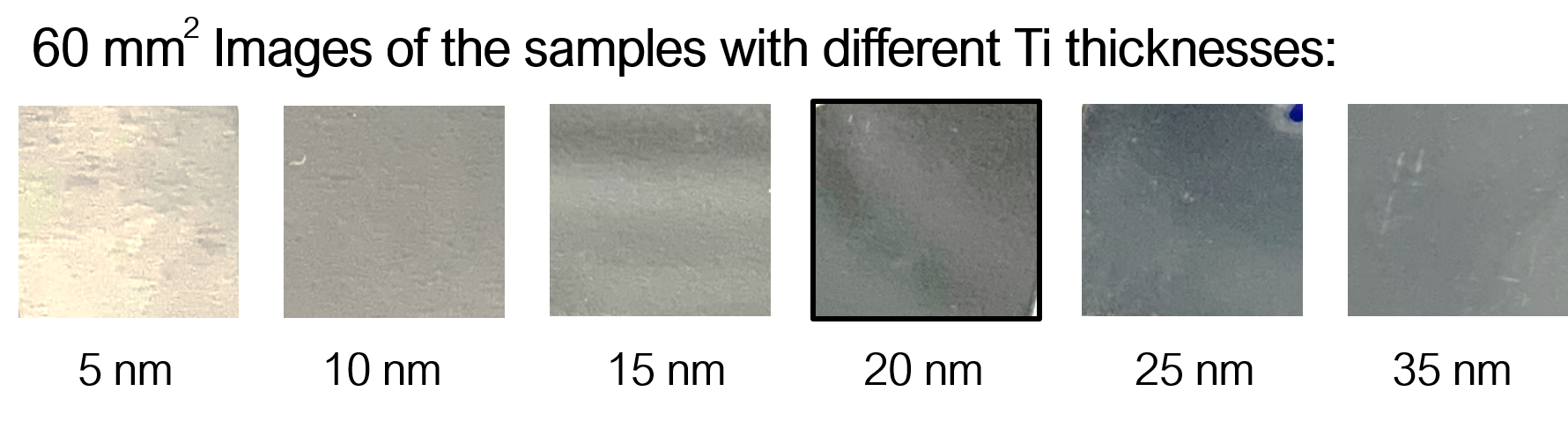} 
    \caption{}
\label{fig:Samples}
    \end{subfigure}\hfill
     \begin{subfigure}{\linewidth}
 \includegraphics[width=\linewidth]{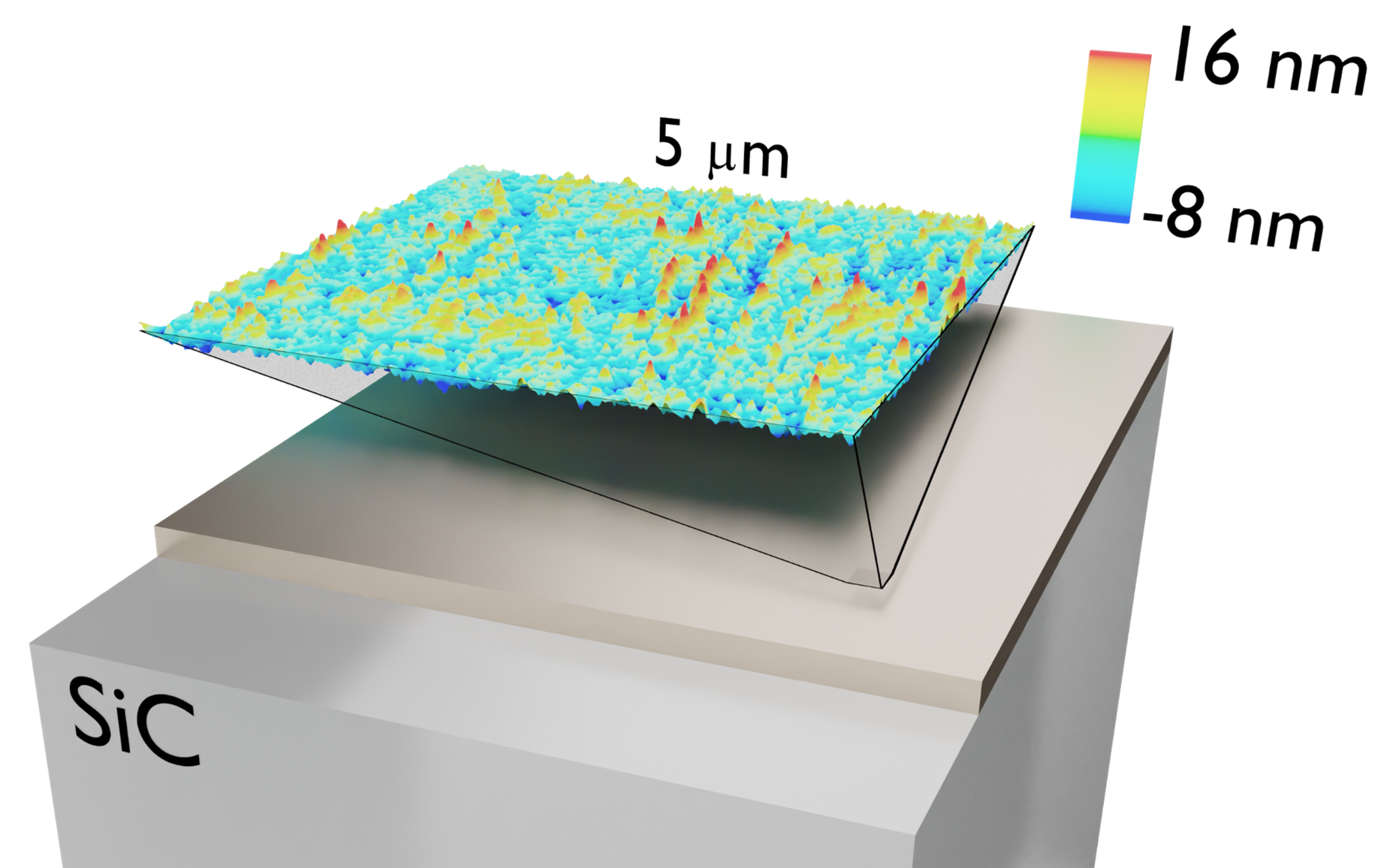}
   \caption{}
\label{fig:Roughness}
    \end{subfigure}\hfill
 \caption{{\bf Fabricated samples.} (a) Photos of the samples with increasing Ti film thickness, deposited on top of a 220 $\mu$m SiC wafer. (b) Detail of a 25 $\mu m^2$ roughness of the selected sample (20 nm Ti thin film) measured via AFM. The RMS roughness is 2.188 nm, mean roughness is 1.635 nm, maximum peak height is 16.05 nm, maximum pit depth is 7.73 nm and the maximum height is 23.78 nm.}
 \label{fig:AFMsample}
\end{figure}

We experimentally demonstrate the spectral stability of material-based HISs by fabricating several samples of titanium (Ti) metallic films deposited on top of a silicon carbide (SiC) substrate, and characterizing them via FTIR spectroscopy. First, we demonstrate the robustness against variations of the thickness of the metallic film by fabricating six samples with different Ti film thicknesses (5 nm, 10 nm, 15 nm, 20 nm, 25 nm and 35 nm) deposited on top of 4H-SiC substrates with 220\,$\mu$m thickness via e-Beam evaporation. Photographs of the fabricated samples are reported in Fig.\,\ref{fig:Samples}, showing that the samples become more opaque in the visible as the Ti film thickness increases. Fig.\,\ref{fig:Roughness} reports the atomic force microscopy (AFM) scan for the sample of 20 nm thickness, with an average roughness of 1.63\,nm. 

\begin{figure}[h]
 \centering
  \includegraphics[width=0.8\linewidth]{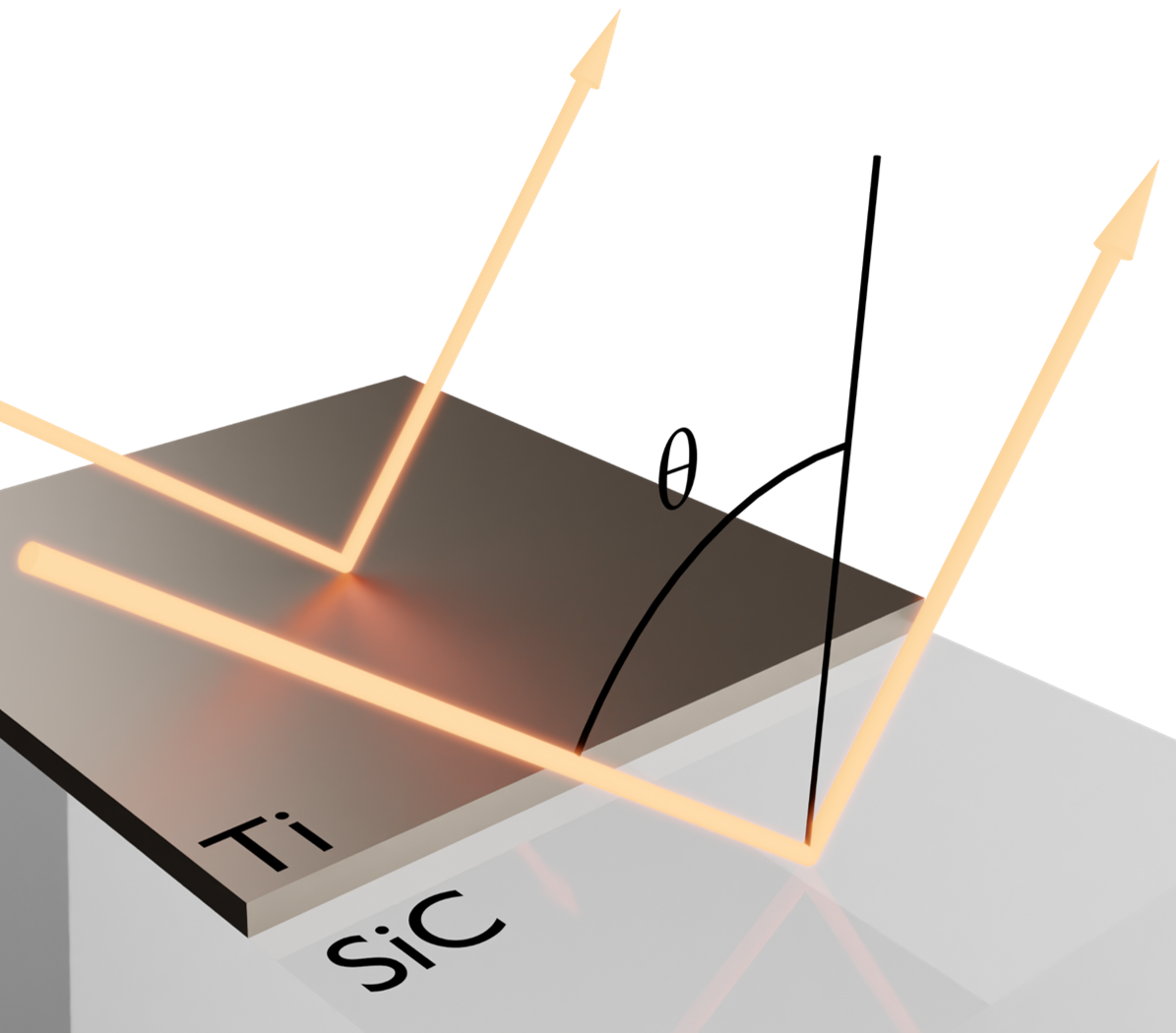}
 \caption{\textbf{Reflectivitty difference}. Representation of reflectivities involved in the substraction: thin film plus substrate ($R_{TiSiC}$) and the substrate itself ($R_{SiC}$), for a determined incidence angle ($\theta$).}
 \label{fig:Reflectsic}
\end{figure}

We characterize the spectral response of the samples via FTIR spectroscopy (see Methods). Unfortunately, the film absorptivity/emissivity cannot be directly measured from FTIR reflectivity measurements, which provide the collective response of film plus substrate system. Therefore, the film absorptivity/emissivity was retrieved by measuring the reflectivity difference, i.e., by subtracting from the reflection of a bare SiC substrate the reflection from a SiC substrate with the thin film placed on it: $A_{Ti}\simeq R_{SiC}-R_{TiSiC}$ (see Fig.\,\ref{fig:Reflectsic}). The reflectivity difference provides an accurate estimation of the film absorptivity/emissivity within the high-reflectivity Reststrahlen band of SiC, and a pessimistic estimation of the film absorptivity/emissivity near the ENZ frequency (see Supplementary Fig.4). Therefore, it is a pessimistic estimate of the spectral performance of our devices.

\begin{figure*}[h]
\centering
    \begin{subfigure}{0.20\linewidth}
    \includegraphics[width=\linewidth]{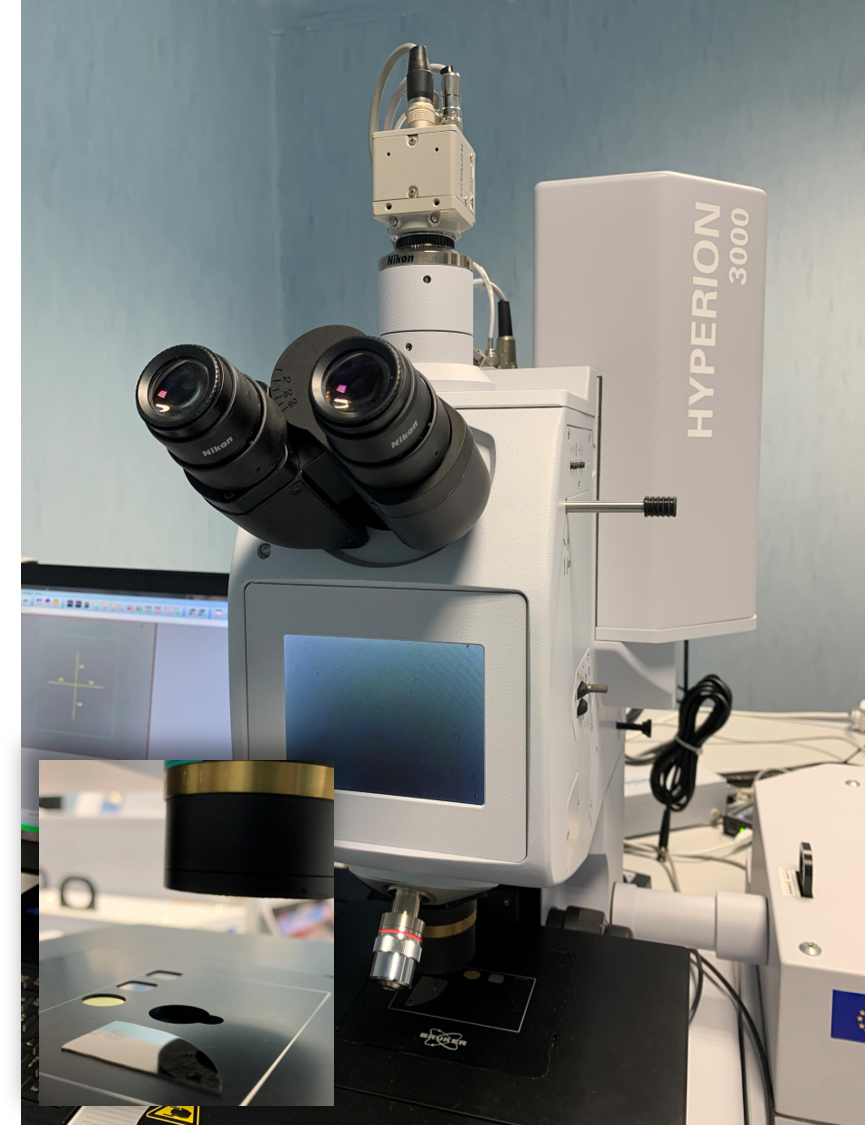}

    \caption{}
\label{fig:NImicroscopio}
    \end{subfigure}
        \begin{subfigure}{0.37\linewidth}
   \includegraphics[width=\linewidth]{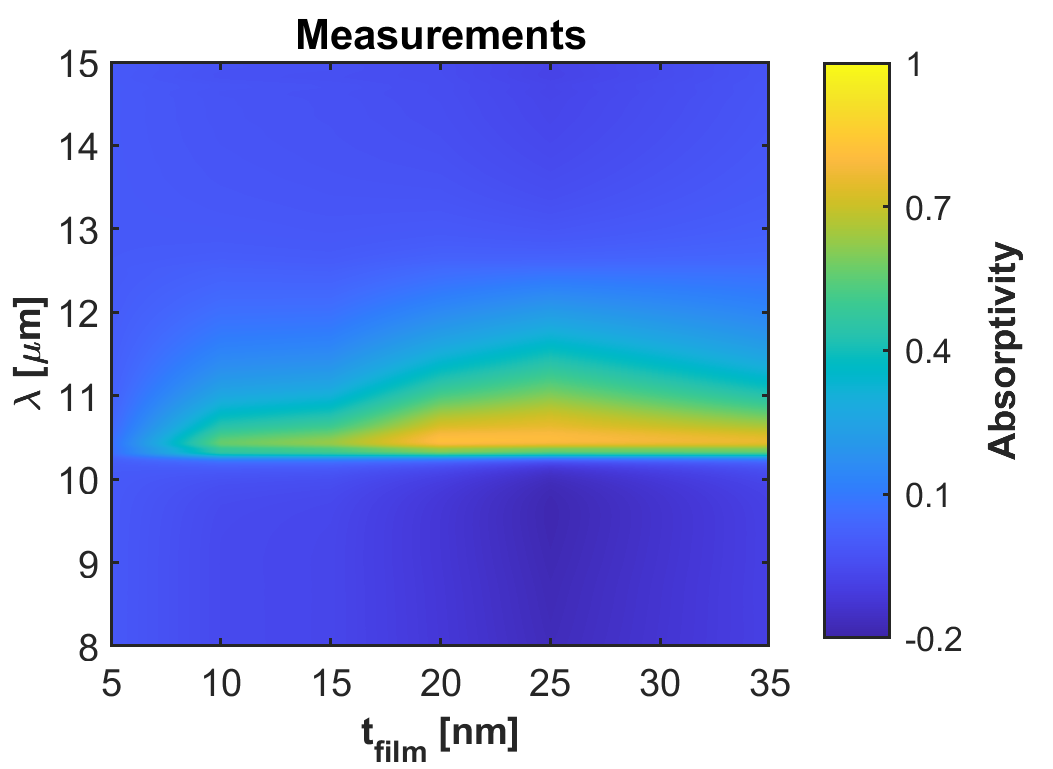}
    \caption{}
\label{fig:NImeasurements}
    \end{subfigure}
    \begin{subfigure}{0.37\linewidth}
    \includegraphics[width=\linewidth]{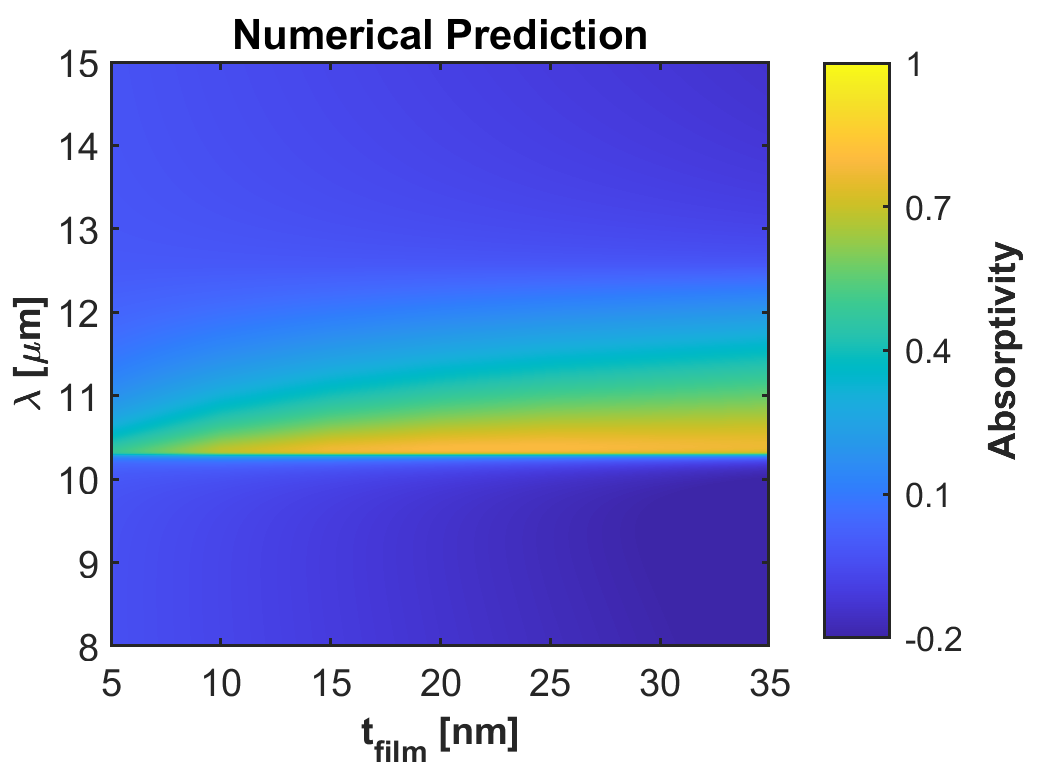}
    \caption{}
\label{fig:NIresults}
    \end{subfigure}
 \caption{{\bf Normal incidence measurements.} (a) Photograph of the FTIR microscope. Comparison between (b) measurements and (c) numerical predictions of the reflectivity difference as a function of the titanium (Ti) film thickness at normal incidence. Measurements were performed on six samples with Ti film thicknesses of 5 nm, 10 nm, 15 nm, 20 nm, 25 nm and 35 nm.}
    \label{fig:NI}
    \end{figure*}

Normal incidence experiments were conducted with unpolarized light and a FTIR infrared microscope (see Fig. \ref{fig:NImicroscopio} and Methods), and the results are reported in Fig.\,\ref{fig:NImeasurements}. The measured reflectivity difference confirms that the peak film absorptivity/emissivity takes place in the vicinity of the plasma frequency of the substrate for all fabricated samples. The reflectivity difference is maximized for a film thickness of 20\,nm, with a peak value of 81.5\,\%. Therefore, the measured sample maximizes the absorptivity/emissivity for a film thickness larger than that predicted by the theory, and it also provides a higher peak value. We ascribe this behavior to the surface roughness of the film, which effectively increases the resistivity of the metallic films. In order to test this hypothesis, we model the rough Ti layer with an effective Drude model: $\varepsilon_{\rm film}=1-\omega_p^2/(\omega(\omega+i\omega_c))$ with fitted values $\omega_p = 3.93\,$x$\,10^{15}$\,rad/s and $\omega_c = 1.06\,$x$\,10^{15}$ rad/s. Numerical predictions based on this fitted lossier Drude model match well with the experimental results as shown in Fig.\,\ref{fig:NI}.

The fact that the experimental data can be explained with a material with increased losses with respect to the theory, supports the conclusion that the fabricated films have a more resistive character. The only major difference between measurements and updated theory takes place for the film of 5\,nm thickness, for which the measured reflectivity is much smaller than even that predicted by the corrected model (see Fig.\,\ref{fig:NImeasurements}, and Supplementary Fig.\,4a). This effect can be justified by the fact that for this sample the surface roughness is comparable to the film thickness. However, the fact the absorptivity/emissivity spectrum is still characterized by a peak at the plasma frequency even for a low quality film, with a roughness comparable to its thickness, supports the conclusion that material-based HISs support an stable spectral response, quite independent of the geometry of the metallic films deposited on top of it.

\begin{figure*}[h]
\centering
    \begin{subfigure}{0.20\linewidth}
    \includegraphics[width=1\linewidth]{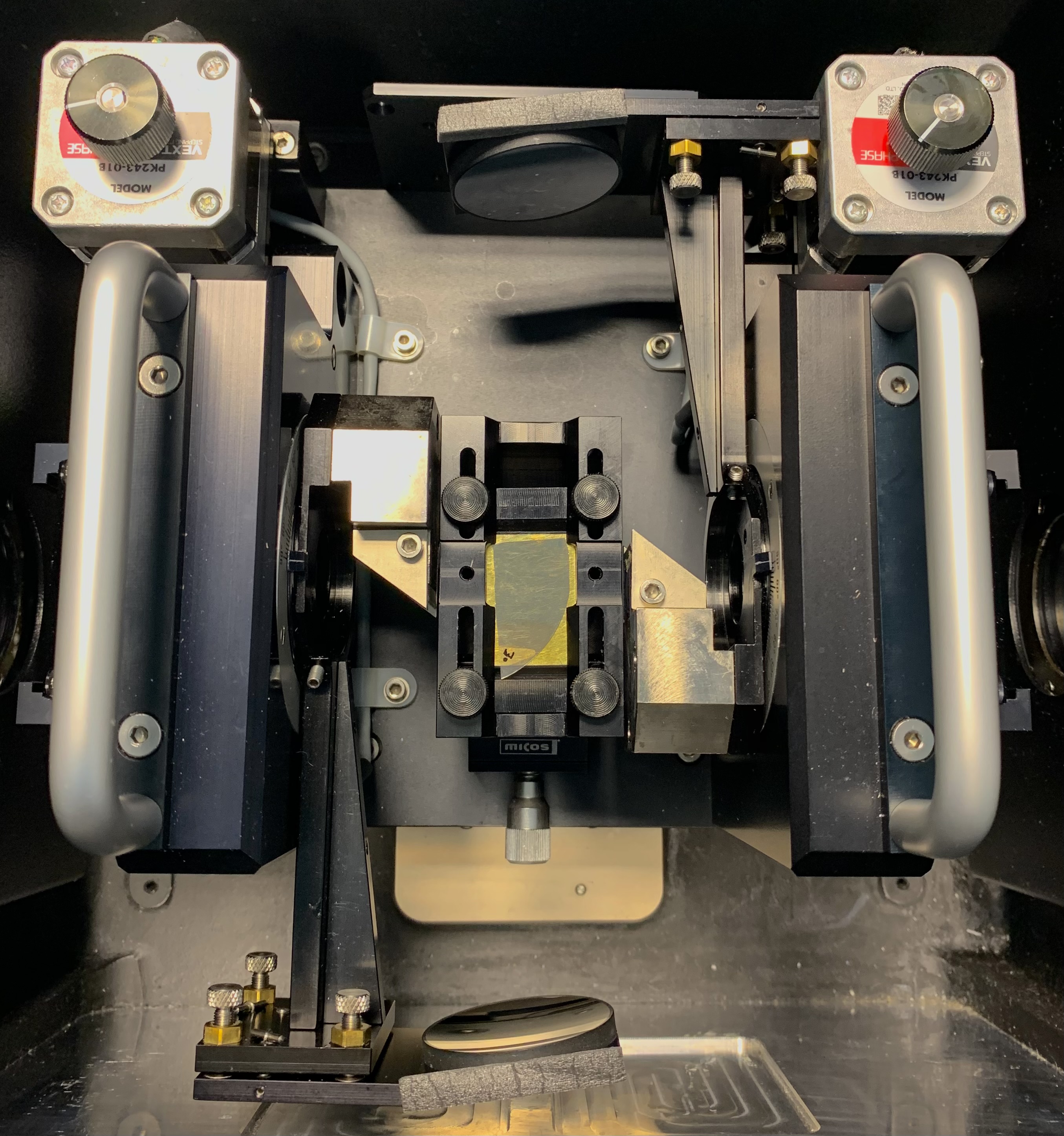}
    \caption{}
\label{fig:OIbruker}
    \end{subfigure}
    \begin{subfigure}{0.37\linewidth}
       \includegraphics[width=1\linewidth]{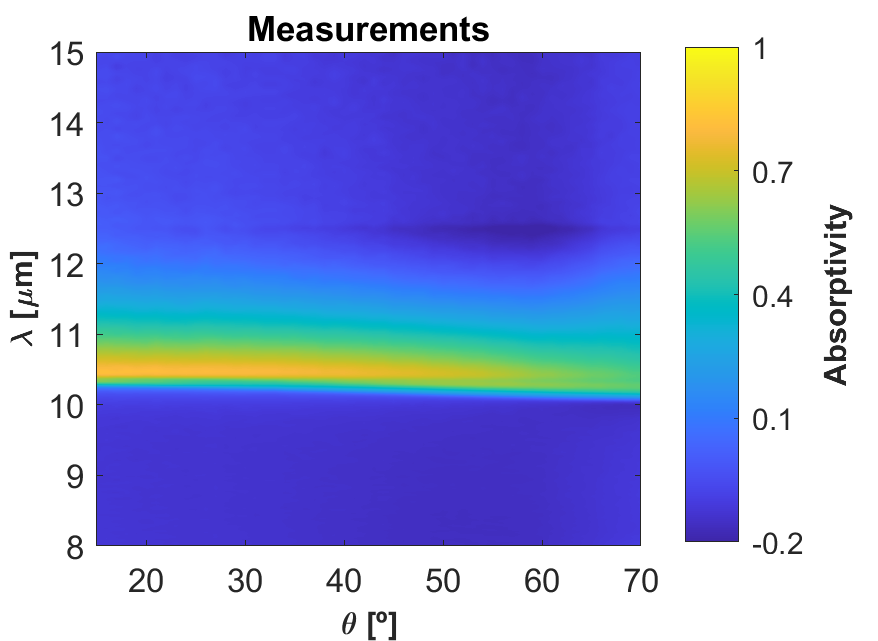}
\caption{}
\label{fig:measurementsanglesweep}
    \end{subfigure}
    \begin{subfigure}{0.37\linewidth}
    \includegraphics[width=\linewidth]{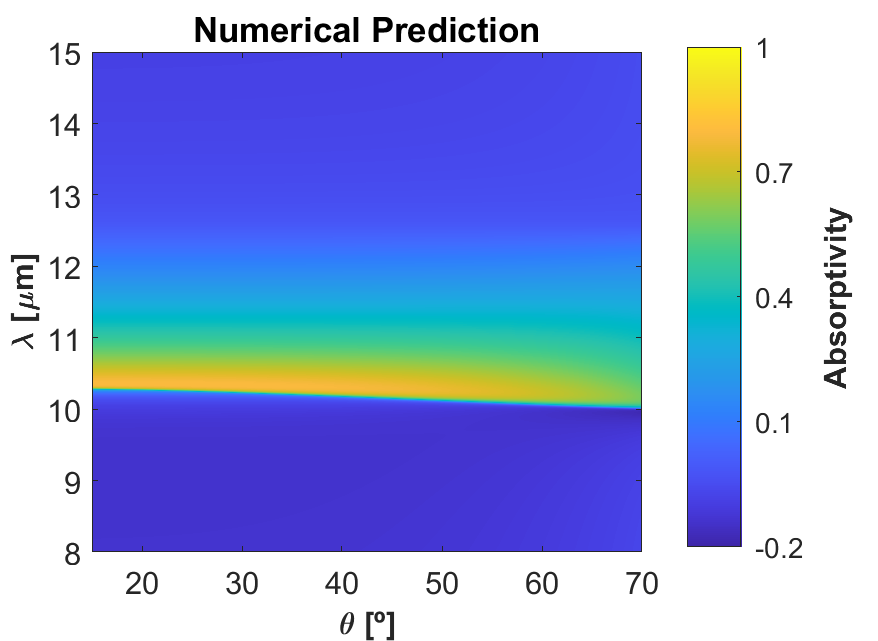}
    \caption{}
\label{fig:expectedresultsanglesweep}
    \end{subfigure}
 \caption{{\bf Oblique incidence measurements.} (a) Photograph of the moving mirror setup for adjusting the incident angle. Comparison between (b) measurements and (c) numerical predictions of the film absorptivity as a function of the incidence angle, for the sample of 20 nm film thickness. Measurements were performed from 15\textdegree\,to 80\textdegree\,with a step of 1\textdegree. }
    \label{fig:expangsweep}
    \end{figure*}

Following the theory, we next analyze the angular response of the 20\,nm thick sample that maximized the absorptivity at normal incidence. To this end, we used a two movable mirror system (see Fig.\,\ref{fig:OIbruker}, Methods) that allows for measuring the reflectance at variable angle of incidence. Measurement results, between 15\textdegree\,and 80\textdegree, with a step of 1\textdegree, are reported in Fig.\,\ref{fig:measurementsanglesweep} as compared to the theory in Fig.\,\ref{fig:expectedresultsanglesweep}. The results demonstrate that the absorptivity/emissivity spectrum of material-based HISs is robust against variations of the observation angle. Moreover, the theoretical comparison of the film absorptivity and the reflectivity difference as a function of the observation angle (see Supplementary Fig.\,5), reveal that the reflectivity difference is a pesimistic estimation of the angular robustness of the film absorptivity/emissivity.

\section{Conclusion}

Our results numerically and experimentally demonstrate that material-based HISs, implemented via ENZ substrates, exhibit an unusually stable and narrowband emission spectrum. By contrast with conventional partially coherent emitters, it was demonstrated that the peak emission frequency is very robust against variations on the system geometry (i.e., the thickness of the metallic film and/or the presence of roughness), the angle of observation, and polarization. In addition, this unique property is achieved without the use of complex nanofabrication processes, opening the pathway to large-area, large-scale and low-cost applications. Our study has focused on samples composed of Ti metallic layers deposited on top of SiC substrates. However, we expect that similar conclusions hold for other material choices both for the metallic films (e.g., Chromium, platinum), and ENZ substrate (e.g., AZO, CdO, ITO). We believe that our study reveals a new class of emitters, which might be of interest for a variety of technologies build around of heat and energy radiative management. Beyond thermal emitters, the absorption of infrared radiation by ultra-thin metallic films is the basis of optoelectronic devices, photothermal modulation and basic research on atomically-thin materials, for which a narrowband emitter with an exceptionally stable spectrum might also be of interest. 

\section{Methods}

\subsection{Simulations}

 Numerical calculations of the film absorptivity/emissivity were carried out with the full-wave solver Comsol Multiphysics \texttrademark \cite{COMSOL}. In particular with the \textit{frequency domain solver} of the \textit{Wave Optics module}. The film is represented as a \textit{Transition Boundary Condition} and its absorptivity is calculated via integration of the dielectric losses. Limits of the unit cell were defined with \textit{Periodic Boundary Conditions}. In the case of the SiC substrate, the bottom limit is presented as \textit{Scattering Boundary Condition} and in the case of a Salisbury Screen as a \textit{Perfect Electric Conductor}. 
 
 Silicon Carbide was modeled with a Lorentzian permittivity \cite{Caldwell2015low}, $\varepsilon_{SiC}(\omega) = \varepsilon_{\infty}(\omega^2 - \omega_p^2 + i\omega\omega_c)/(\omega^2 - \omega_0^2 + i\omega\omega_c)$, with normalized medium impedance $Z_{SiC}(\omega)=1/\sqrt{\varepsilon_{SiC}(\omega)}$. 
 In the Theory and Numerical Results Section, Titanium was modeled with tabulated data \cite{Rakic98}, in the Experimental Results section, it was modelled with an effective Drude model: $\varepsilon_{\rm film}=1-\omega_p^2/(\omega(\omega+i\omega_c))$ with fitted values $\omega_p = 3.93\,$x$\,10^{15}$\,rad/s and $\omega_c = 1.06\,$x$\,10^{15}$ rad/s. 
 
\subsection{Samples fabrication and measurement}
  
The samples were fabricated in the ISO7 cleanroom at UPNA facilities evaporating Ti on 4H-SiC wafer with an Angstrom Engineering e-Beam evaporator. The six samples are shown in the Fig.\,\ref{fig:Samples}. Normal incidence measurements of the samples reflection were performed via Fourier Transform Infrared Spectroscopy (FTIR) with a microscope Hyperion 3000, and the oblique incidence measurements with the coupled accessory A513/Q in the Bruker Vertex 80v spectrometer.

\section{Acknowledgements}

I.L. acknowledges support from Ram\'on y Cajal fellowship RYC2018-024123-I and project RTI2018-093714-301J-I00 sponsored by MCIU/AEI/FEDER/UE, and
ERC Starting Grant 948504. We thank Mar\'ia Elena Hern\'andez-Mart\'inez for her support with AFM scans. 

\bibliography{library.bib}
\bibliographystyle{IEEEtran}
\end{document}